# Prediction Approach against DDoS Attack based on Machine Learning Multiclassfier

Anupama Mishra
Department of Computer Science
Engineering,
Swami Rama Himalayan University
tiwari.anupama@gmail.com

*Abstract*— **DDoS attacks, also known as distributed denial of service (DDoS) attacks, have emerged as one of the most serious and fastest-growing threats on the Internet. Denial-of-service (DDoS) attacks are an example of cyber attacks that target a specific system or network in an attempt to render it inaccessible or unusable for a period of time. As a result, improving the detection of diverse types of DDoS cyber threats with better algorithms and higher accuracy while keeping the computational cost under control has become the most significant component of detecting DDoS cyber threats. In order to properly defend the targeted network or system, it is critical to first determine the sort of DDoS assault that has been launched against it. A number of ensemble classification techniques are presented in this paper, which combine the performance of various algorithms. They are then compared to existing Machine Learning Algorithms in terms of their effectiveness in detecting different types of DDoS attacks using accuracy, F1 scores, and ROC curves. The results shows high accuracy and good performance.**

## 1. Introduction

DDoS (distributed denial-of-service) attack originates from many sources scattered over multiple network locations. DoS attacks are primarily motivated by the desire to significantly degrade the performance or completely consume a certain resource, and a process to exploit a machine defect and cause failure of a processing or exhausting the system resources by exploiting a system flaw. Yet another method of assaulting the target system is to flood the network and monopolise it, so preventing anyone else from utilising it [1]. DoS attacks are defined and classified by the prohibition of access to the victim machine or network, whereas DDoS attack is the use of a large number of systems from distributed environment to launch the attack, which is defined and classified by the use of many computer systems or services. Keep in mind that attack agents can be any vulnerable devices or resource that has the capability of running the suspicious code, such as Internet of Things devices, networked PCs, servers, and armed mobile devices, among other things [2].

Figure 1 shows that Telecommunication industry was affected severely by this attack[3]. Also it can be seen from the figure 2 that the united states is highly suffered country by this attack[3].

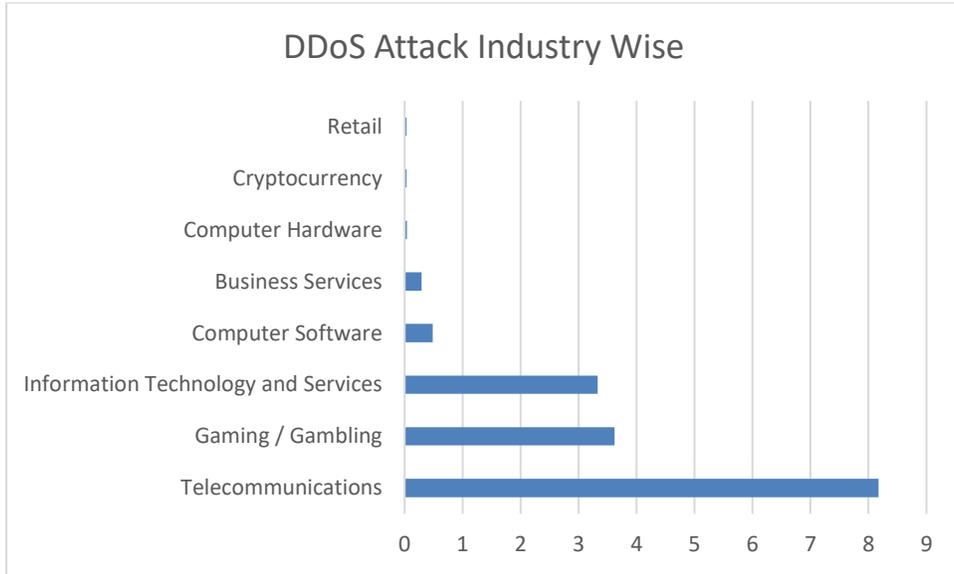

Figure 1: Statistics of DDoS attacks Industry wise

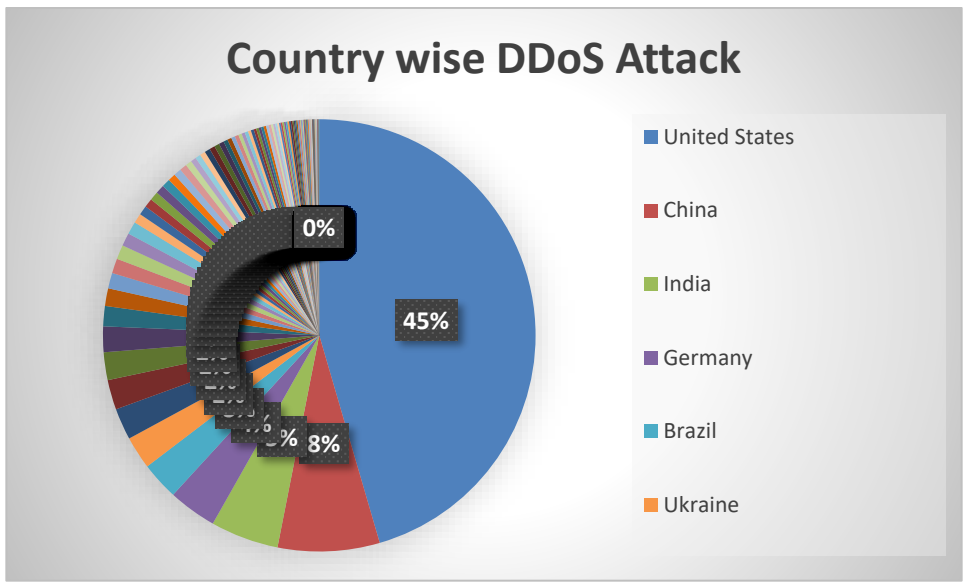

Figure 2: DDoS attack Country wise

There are numerous defensive measures available against distributed denial of service (DDoS) assaults[4-5], but none of them is completely effective. There are a variety of factors contributing to this, including the following:

- Everyone has the ability to launch an attack due to the quantity of open-source tools available on the internet.

- DDoS attacks almost always include forged IP packets, making it nearly impossible to identify the source of an attack.
- DDoS attacks are becoming increasingly sophisticated. In addition, the length of an attack has dropped in recent years to around 4 minutes, down from previously. The affected machine crashes as a result, preventing any defence solution from detecting the attack. As a result, acquiring complete information on distributed denial of service (DDoS) assaults is extremely difficult.
- It is impossible to directly compare defensive devices with their competitors on the market because there is no common benchmark for DDoS defence filters in the computer sector.
- Cloud computing, fog computing, industrial computer systems and the Internet of Things are among the many innovative technologies that are making their way to market.

Improve standard defence methods so that they may be used in these conditions is a difficult and time-consuming task that will need a great deal of effort to do.

In the following sections, you will find an outline of the paper: A discussion and analysis of related existing studies for protective methods against distributed denial of service are presented in Section 2. (DDoS). Section 3 introduced and examined our method, as well as the fundamental concepts that drive it. Section 4 introduced and explored our approach. In Section 4, we provide an evaluation of our proposed work, and Section 5 of the paper finishes the research with a conclusion.

## 2. Literature Review

The DDoS attack is not a new attack but even though after having so many existing techniques, the attackers are able to perform this attack. The reason behind it is lack of security concern, the existing mechanisms are having limitations and exploration of the vulnerabilities in terms of parameters used in network. Several security strategies were proposed [5-8] for the identification and detection of distributed denial-of-service attacks (DDoS) in this study.

It was advocated by the authors in [6] that a low-cost defensive method against DDoS attacks should be based on differences in entropy between DDoS assaults and normal traffic be used. In addition, the authors developed a method for weakening the onslaught's overall strength. It offers several advantages over other current systems, including a high detection rate, a low false-positive rate, and the potential of mitigating the problem. It is currently in development.

Authentication and security issues related with smart vessels in marine transport are addressed by the authors [7] through the development of a method for handling these issues. By authenticating devices in marine transport and detecting various cyberattacks, such as distributed denial of service (DDoS). The proposed method authenticates smart vessel access using an identity-based methodology, which is described in detail below. This strategy, on the other hand, is limited to maritime transportation.

The authors in reference [9] discuss the development of a secure data sharing method as well as a cyber-attack detection methodology that make use of identity-based encryption (IBE) and deep learning algorithms as part of their research. To manage sensitive information, the proposed system makes use of identity-based encryption.

The authors' [10] goal in this study is to provide a better understanding of a comprehensive framework for comprehending the characteristics of vulnerabilities in information systems, such as the vulnerability category a specific vulnerability falls under belongs to, the potential threats it poses, and the warning signs that it is approaching for their assistance in fixing the issue. As an additional source of data, the authors use a leading vulnerability report site to compile information on actual vulnerabilities discovered in companies' information systems.

Authors in [11] suggested a DDoS detection system that was boosted. The method in Internet of Things (IoT) devices RFID tags were proposed by the authors for reciprocal identification. Devices connected to the Internet of Things require authentication.

In 2021, the author [12] proposed a game theory-based security mechanism for analysing data. DDoS attacks are being mitigated by the use of a multi-attribute based auction. In this case, the According to the authors, a reputation-based detection system is developed, in which the, the reputation of a user is determined by his or her marginal utility. There are two separate Various payment methods for both legitimate and fraudulent consumers have been proposed.

Table 1: Comparative Study of Existing Approaches

| Approach | Mechanisms | Contribution | Limitation |
| --- | --- | --- | --- |
| Mishra et al., 2021 | Software Defined Network | High Detection Rate | Not able to identify different type of DDoS. |
| Zhou et al., 2021 | Identity Based Encryption | High Detection Rate | High Complexity |
| Gupta et al., 20220 | Identity Based Signature | Moderate Detection rate | Increase Overhead |

| Cviti´c et al., 2021) | Boosting Based | Moderate Detection rate | Did not reduce false alarm |

## 3 Proposed Methodology

Following a detailed discussion of the dataset, we will apply a preprocessing procedure that includes feature selection. At the end of the process, we train and test the model on datasets that have been selected based on their features.

### 3.1 Approach

In our paper, we are focusing on preprocessing of the data, feature selection[13], model the machine by selecting classifier and then prediction on the unseen data and based on performance evaluation we validate our prediction on new unseen data. The approach is as following:

Preparation of data: The third phase is comprised of a number of tasks that are all focused on turning the raw data gathering into a finished dataset, as described above. The nature and order of activities may differ, and some tasks may even be repeated more than once, depending on the status of the raw data at the time of completion. Data cleansing, feature selection, and data transformation are just a few of the responsibilities involved.

Modelling: When selecting and applying relevant modelling techniques to the data, the fourth phase is called the selection phase. In most cases, the parameters of these models are calibrated in order to get the best possible performance.

This process is strongly related to data preparation since modelling may reveal new issues with the data that were not previously apparent. In addition, the manner in which the data is prepared can result in the usage of a variety of models.

Performance Evaluation: During the evaluation phase, the models that were used in the previous phase are thoroughly assessed and reviewed by experts. During this phase, the tasks completed are compared to the planned objectives in order to confirm that all business requirements have been taken into consideration and are being met. Furthermore, the models are assessed for generalisation to data that has not yet been observed. There needs to be a clear understanding of how the data mining results should be implemented by the time the evaluation step is completed.

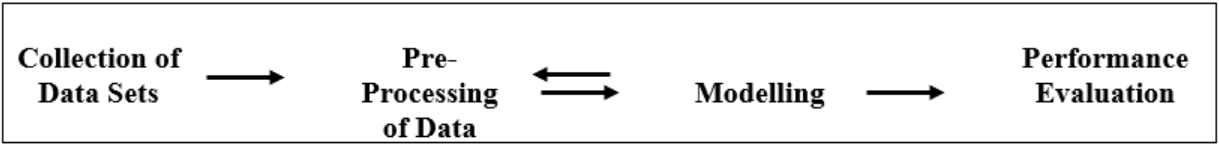

Figure 3: Proposed approach for defending the DDoS attacks.

### 3.2 DataSet and Preprocessing

Datasets are an essential component of machine learning algorithms. Any type of machine learning model conceptualization will be successful if a dataset with real-world circumstances or events to be categorised is obtained or developed and included. During both the training and testing phases of the model, this is true. CICDoS2019 [14] is a dataset supplied by CIC that is designed to offer trustworthy information for the training of DDoS attack detection algorithms.

Many different types of attacks that were implemented in the CIC dataset were divided intoclass label for experiments

It is estimated that the dataset includes sufficient information for model training and validation because it comprises a total of 5775786 rows of information in CSV extension files. There are a total of 88 attributes in the CIC DDoS dataset, which are listed in table 2 below.

Table 2: Shape of the dataset in number of rows and columns

| Total No of Rows | Total No of Columns |
| --- | --- |
| 5775786 | 88 |

Figure 4 presents the Class Label along with its total occurrence in the dataset as follows the MSSQL attack-5763061, LDAP-9931 and BENIGN-2794

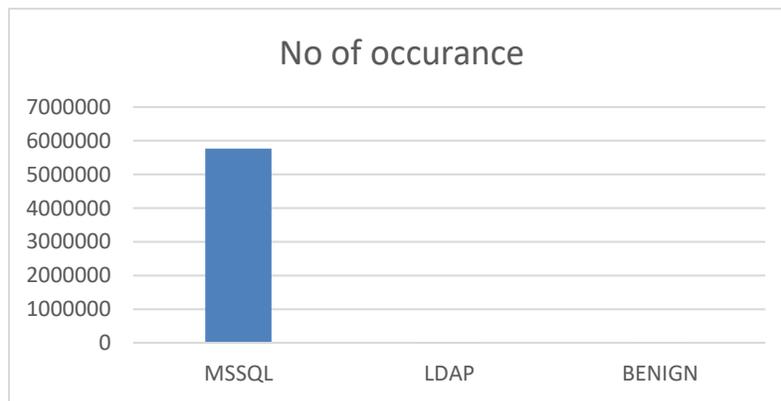

Figure 4: Type of attacks along with its total occurrence in the dataset

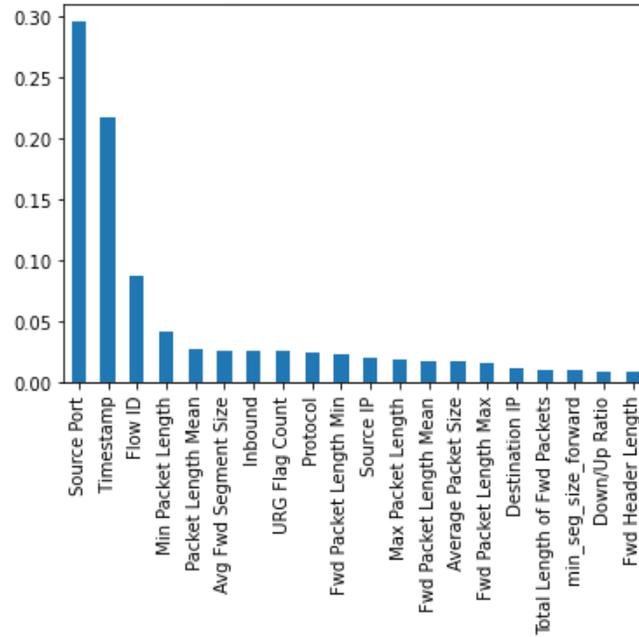

Figure 5: Highest 20 features by using Extra Tree Classifier

In our research, we have use Extra Tree Classifier for selecting the best 20 features as depicted by figure 5. Also, Table 2 shows the selected features and their description.

Table 3: List of selected 20 features and their description

| S.No | Selected Features | Description |
| --- | --- | --- |
| 1. | Timestamp | Time spends during session |
| 2. | Min Packet Length | Minimum length of a packet |
| 3. | Source Port | Source port utilized by the session. |
| 4. | Protocol | Protocol used by Session |
| 5. | Packet Length Mean | Mean length of a packet |
| 6. | Avg Fwd Segment Size | Average size observed in the forward direction |
| 7. | Fwd Packet Length Min | Minimum size of packet in forward direction |
| 8. | Average Packet Size | Average size of packet |
| 9. | Fwd Packet Length Max | Maximum size of packet in forward direction |
| 10. | Max Packet Length | Maximum length of a packet |
| 11. | Fwd Packet Length Mean | Mean size of packet in forward direction |
| 12. | Flow ID | An internal numerical identifier applied to each Flow |
| 13. | Total Length of Fwd Packets | Total size of packet in forward direction |
| 14. | Init_Win_bytes_forward | The total number of bytes sent in initial window in the forward direction |
| 15. | Down/Up Ratio | Download and upload ratio |
| 16. | Inbound | Interface that the session was sourced from. |
| 17. | Destination Port | Destination port utilized by the session |
| 18. | Source IP | Original session source IP address |
| 19. | Fwd Header Length | Length of header in forward direction |
| 20. | Min_seg_size_forward | Minimum size of segment in forward direction |

### 3.3 Modelling

To analyze the datasets, four different supervised learning models have been chosen for comparison. Several criteria[15-17] are used to select the models, which include the presence of both parametric and nonparametric models, the use of algorithms from a number of categories, and the application of models that have been extensively used in previous research and publications.

### 3.3.1 Random Forest

The random forest algorithm is an ensemble technique that classifies data by using a large number of decision trees in a random fashion. Ensemble algorithms[18-20] are more accurate than single model algorithms because they combine numerous models. Each decision tree is randomly selected from a training set, and then each decision tree's vote is aggregated from all of the other decision trees in a random fashion, and the final object examined is provided. A primary reason for using the classifier is because it performs well with huge datasets and can handle a large number of input variables.

### 3.3.2 Decision Tree

The classification process in a decision tree begins at the root node and categorises observations based on the values of the characteristics associated with each observation. Nodes indicate a single characteristic, and the values that they can adopt are represented by their corresponding nodes[21-24].

Working its way down from the root node, the algorithm iteratively computes the information gain for each feature in the training set, starting with the most significant feature. To estimate the extent of discrimination imposed by the features towards the target groups, it is necessary to consider information gain. When it comes to categorising each observation, the higher the information acquisition, the greater the value of the attribute. The root node is replaced by the attribute that provides the greatest amount of information gain, and the algorithm continues splitting the data set by the selected feature to produce subsets of the original data set.

### 3.3.3 SVM

It is a supervised learning approach; it divides the different classes using a hyper plane and then constructs a model that is capable of detecting previously unseen samples[25-27]. For multilabel classification, LinearSVC from Sci-Kit Learn was used using the parameter "ovr" one-vs-rest, which stands for one-versus-rest. The square hinge was picked as the loss function because it produces results that are both computationally efficient and effective. In order to lower the margin, the regularisation and cost parameters are both set to one.

### 3.3.4 Naïve Bayes

This classifier is based on Bayes' Theorem, which assumes that all events are independent of one another. In statistics, two events are said to be independent if the probability of one does not have an impact on the probability of the other[28].

## 4. Experimental Results and Performance Analysis

For the evaluation of the performance in classification machine learning[29-31], we have the following metrics:

Accuracy: It is defined as correctly classified from the total samples.

Recall: The recall rate is a measure of how many of the real positives were discovered or recalled after the fact.

Precision: Precision can be defined as the proportion of accurately identified positives, also known as true positives, in a sample.

F-Score: In machine learning, the F-measure, which is a combination of precision and recall, provides an overall accuracy score for a model. A good F-measure score indicates that a model has low false positives as well as low false negatives, and as a result, a model that correctly identifies threats while generating the fewest false alarms is preferred.

RoC Curve: The Receiver Operating Characteristic (ROC) curve is used to evaluate the performance of a classification model by taking into account the False Positive Rate (FPR) and True Positive Rate (TPR).

Table 4 and figure 6 shows the values of Recall, Precesion and F Score and Table 5 and figure 7 shows the accuracy results of all the applied classifiers. Also Figure 8,9,10, and 11 are used to present the RoC graph of RANDOM FOREST, DECESION TREE, NAÏVE BAYES, AND SVM.

Table 4: Performance Metrices of Selected Machine learning Classifiers

|  | Random Forest | | | Decision Tree | | | Naïve Bayes | | | SVM | | |
| --- | --- | --- | --- | --- | --- | --- | --- | --- | --- | --- | --- | --- |
|  | Precesion | Recall | F-1 Score | Precesion | Recall | F-1 Score | Precesion | Recall | F-1 Score | Precesion | Recall | F-1 Score |
| BENIGN | 100 | 100 | 100 | 40 | 50 | 50 | 77 | 100 | 87 | 100 | 100 | 100 |
| DDoS_LDAP | 100 | 100 | 100 | 97 | 99 | 98 | 98 | 99 | 99 | 99 | 99 | 99 |
| DDoS_MSSQL | 100 | 100 | 100 | 100 | 100 | 100 | 100 | 100 | 100 | 100 | 100 | 100 |

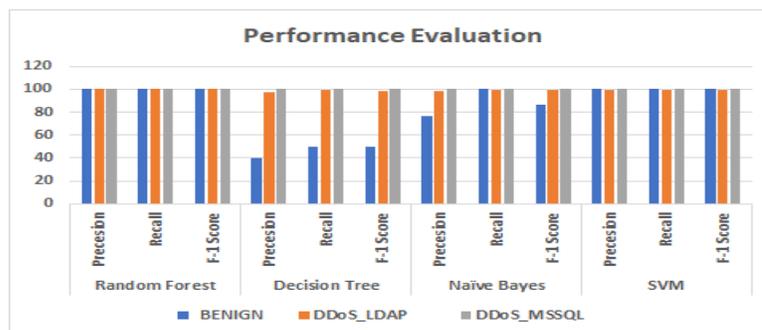

Figure 6: Performance Metrices of applied machine learning classifier

Table 5: Accuracy of used algorithms

| Algorithms | Accuracy |
|---|---|
| **Random Forest** | **99.99** |
| **Decision Tree** | **99.89** |
| **Naïve Bayes** | **99.98** |
| **SVM** | **99.99** |

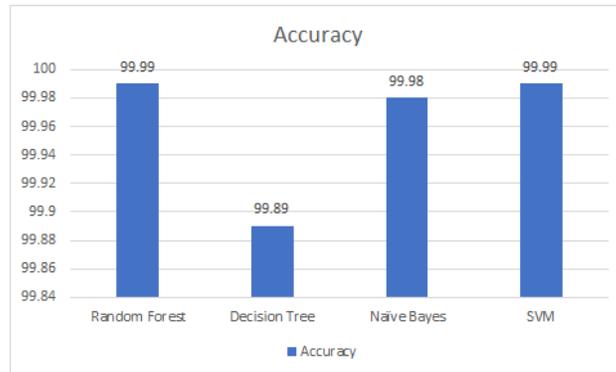

Figure 7: Accuracy Results of applied machine learning classifier

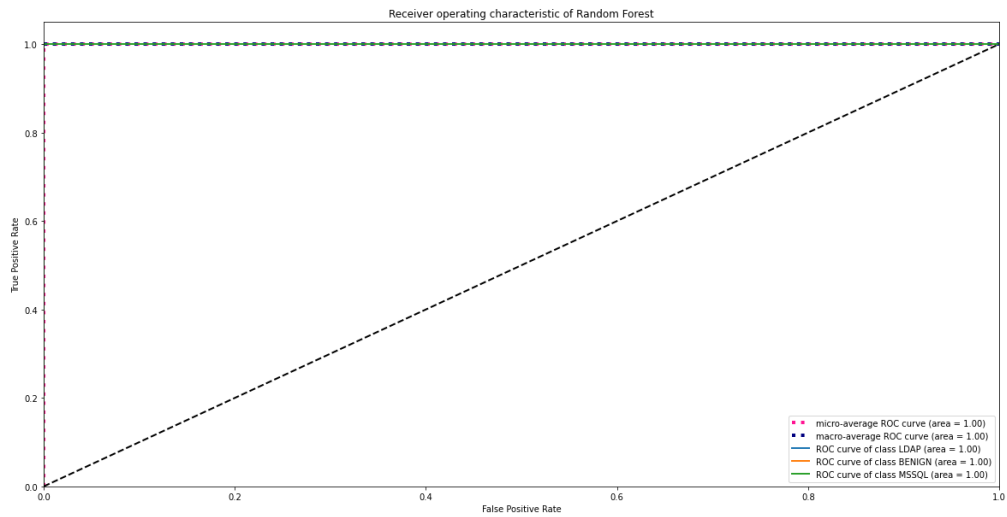

Figure 8: RoC Curve of RANDOM FOREST

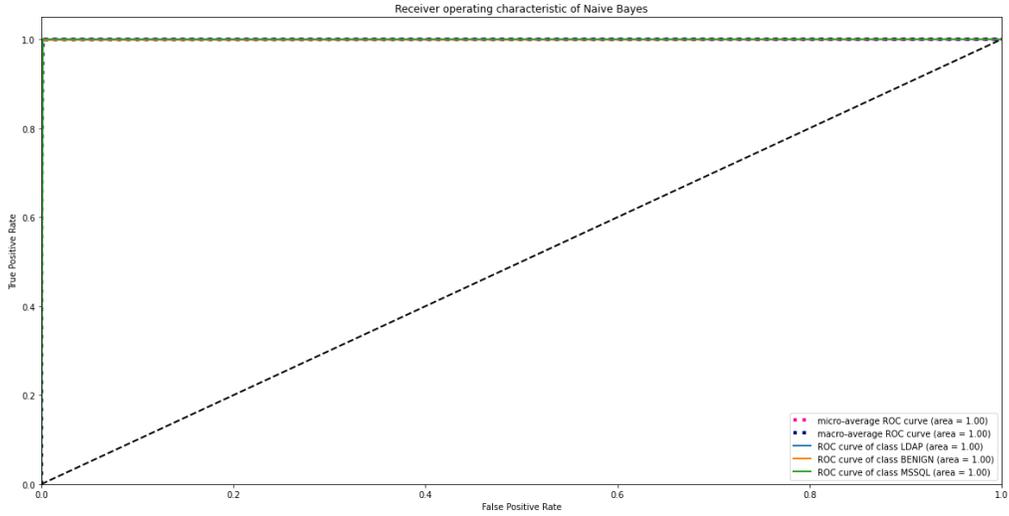

Figure 9: RoC Curve of NAÏVE BAYES

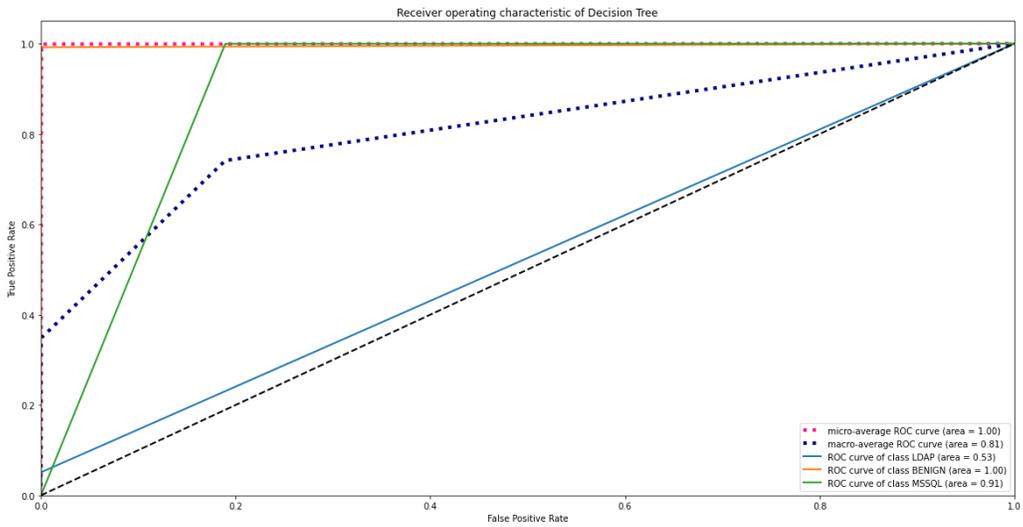

Figure 10: RoC Curve of DECISION TREE

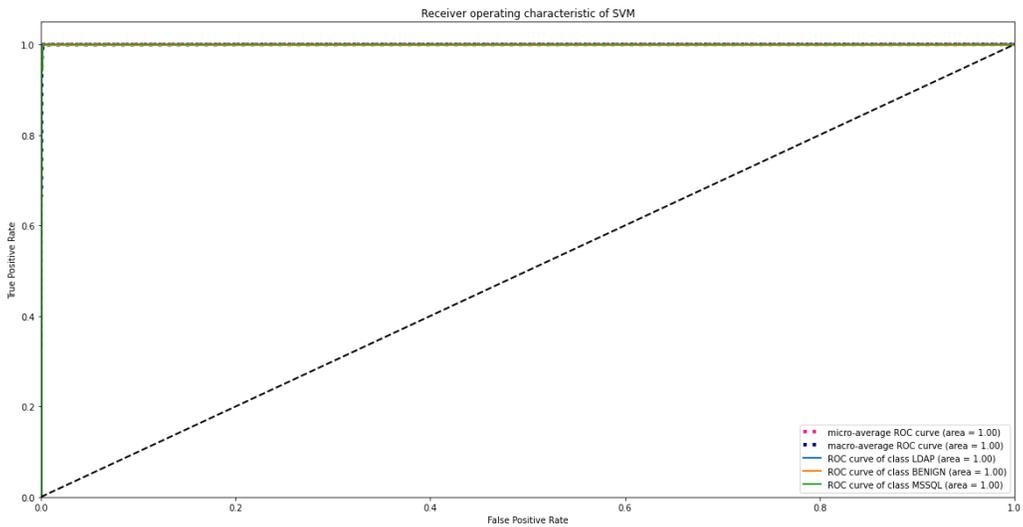

Figure 11: RoC Curve of SVM

In this section, we discuss and analyze the results obtained from each of the algorithms on our processed dataset and depicted in figures from 6 to 11. As per table 4. All four performed well in classified the MSSQL traffic and both are able to identify benign traffic 100%. But in case of LDAP Random Forest Performed well in comparison to other three classifiers, however decision tree and naïve bayes are some how failed to recognize the benign traffic properly.

By the analysis of RoC Curve also , the result of the performance metrices can be verified.

## 5. Conclusions and Future Work

The DDoS attack was classified into multiple classes using ML algorithms, and each type was detected and validated using distinct criteria. For DDoS multiclass Cyberthreat identification, a comprehensive analysis of multiple ML algorithms was conducted and Random Forest and SVM Classifier having the highest accuracy score of 99.99%. However, the naïve bayes got the 99.98% and decision tree 99.89% accuracy to achieve the target. Here we have targeted 3 class labels such as benign traffic, MSSQL and LDAP traffic, in future many other types of DDoS attack can also be targeted for classification and prediction.